\documentclass[pra,%
 reprint,
 amsmath,amssymb,
 aps,
]{revtex4-2}

\usepackage{graphicx}
\usepackage{dcolumn}
\usepackage{bm}

\usepackage{xcolor}
\usepackage[colorlinks,citecolor=blue,linkcolor=blue,urlcolor=blue,hyperindex]{hyperref}

\begin{document}

\preprint{APS/123-QED}

\title{SU(2) symmetry of spatiotemporal Gaussian modes propagating in the isotropic dispersive media}

\author{Fangqing Tang$^{1,2}$}
\author{Xing Xiao$^1$}
\email{xiaoxing@gnnu.edu.cn}
\author{Lixiang Chen$^2$}%
\email{chenlx@xmu.edu.cn}
\affiliation{$^1$School of Physics and Electronic Information, Gannan Normal University, Ganzhou, Jiangxi 341000, China
}
\affiliation{$^2$Department of Physics, Xiamen University, Xiamen, Fujian 361005, China
}

\date{\today}

\begin{abstract}
The far-field intensity distribution of spatiotemporal Laguerre-Gaussian (STLG) modes propagating in free space exhibits a multi-petal pattern analogous to that observed in tilted Hermite-Gaussian modes. Here, we show that this phenomenon can be explained by the SU(2) symmetry of spatiotemporal Gaussian modes, which can support the irreducible representation of SU(2) group and enable the construction of the spatiotemporal model Poincaré sphere. We have also derived analytical expressions for the STLG mode with an arbitrary radial and angular indices propagating in the isotropic media. The propagation dynamics can be understood as a unitary transformation generated by a conserved quantity, where the rotation angle is exactly the intermodal Gouy phase of the spatiotemporal modes in the same order subspace. This spatiotemporal Gouy phase depends on the ellipticity of the wave packets and the group velocity dispersion (GVD) of the media. The phase, as a function of propagation distance, is categorized into three distinct regimes: normal dispersion, anomalous dispersion, and zero dispersion. Interestingly, in the regime of anomalous dispersion, the non-monotonic behavior induces to both distortion and revival of the intensity distribution, thereby establishing a phase-locked mechanism that is analogous to the Talbot effect.
\end{abstract}

\maketitle


\section{\label{sec:level1}Introduction}

Spatiotemporal optical vortices (STOVs) represent a class of polychromatic pulses whose spiral wavefronts resides in spatiotemporal planes (parallel to the propagation direction) and carry transverse orbital angular momentum (TOAM)~\cite{wan2023optical}. Recent studies have demonstrated that the intensity distribution of STOVs would change dramatically when propagating in the free-space ~\cite{chong2020generation,hancock2019free}, in contrast to the self-similar transformation of conventional spatial vortex beams. This unique propagation feature of STOVs stems from the difference between frequency dispersion and spatial dispersion of the media. In 2019, Hancock \textit{et al}. have systematically studied the propagation feature of STOVs in free-space, observing that a STOVs with topological charge 1 would be split into 2 lobes at far field~\cite{hancock2021mode}. Subsequent studies by Huang \textit{et al}. extended the observation to higher-order modes, showing that a vortex with a topological charge $l$ would be split into $|l|+1$ lobes~\cite{huang2022diffraction}. Recently, Vo \textit{et al}. have derived rigorously the closed-form solutions for the STOVs propagating in free space or dispersion-free media, where the solutions satisfy the Helmholtz equation and the splitting behavior can be interpreted by a ray-optics model~\cite{vo2024closed}.

For quasi-monochromatic and paraxial optical pulses, their slowly varying amplitude in dispersive media obeys a Schrödinger-type propagation equation and the frequency dispersion is governed by the group velocity dispersion (GVD). The spatiotemporal Laguerre-Gaussian (STLG) modes~\cite{liu2024spatiotemporal}, as a most representative type of STOVs, is the generalization of conventional Laguerre-Gaussian beams into spatiotemporal plane. In similar, the STLG modes with arbitrary radial index ($p$) and azimuthal index ($l$) can also form a complete orthogonal basis in spatiotemporal plane. The work by Hancock \textit{et al}. developed a comprehensive framework for spatiotemporal Hermite-Gaussian (STHG) modes propagating in isotropic dispersive media~\cite{hancock2021mode}, but their results were limited to the first-order STLG modes. In 2023, Porras \textit{et al}. derived propagation solutions for high-order STLG modes with $p=0$ and arbitrary $l$ in free space~\cite{porras2023propagation} and later extended their solutions in the isotropic media with linear and second-order dispersion~\cite{hyde2023propagation}. Simultaneously, Bekshaev developed another general approach to derive the propagation~\cite{bekshaev2024spatiotemporal}. The interpretation of the splitting behavior is that the absence of energy flow along the temporal direction prevents the maintenance of cylindrical intensity distribution~\cite{porras2024procedure}. However, to the best of our knowledge, no analytical expressions for high-order STLG modes with arbitrary $p$ and arbitrary $l$ have been reported, which necessitates further investigation. Here, we employ SU(2) group formalism to derive the analytical expressions of the STLG modes with arbitrary $p$ and arbitrary $l$ propagating in the isotropic media with dispersion. Furthermore, we elucidate the underlying mechanism responsible for the splitting behavior.

We notice that the splitting behavior of STOV closely resembles the mechanism that a conventional Laguerre-Gaussian (LG) beam is converted into a 45°-tilted Hermite-Gaussian (HG) beam when passing through a mode converter composed of a pair of cylindrical lenses~\cite{beijersbergen1993astigmatic}, where the transformation of STOVs occurs solely through free-space propagation without the mode converter. From the perspective of group action theory, this transformation corresponds to a SU(2) ‘rotation’ operation. Unlike intuitive spatial rotations in real three-dimensional space, this SU(2) ‘rotation’ manifests as a rotation of the Wigner distribution function in phase space~\cite{calvo2005wigner}, that is also referred to as the antisymmetric fractional Fourier transform~\cite{gutierrez2020modal}. A recent study by Bekshaev has demonstrated that the Wigner distribution function and intensity moments serve as effective tools for analyzing the evolution of intensity and TOAM in STOVs propagating through ABCD systems~\cite{bekshaev2024wigner}. However, the underlying SU(2) symmetry warrants further investigation. Given that LG and HG modes represent two bases capable of spanning the invariant subspaces of the irreducible representations of SU(2)~\cite{dennis2017swings}, it is important to note that rotation operations do not change the total order $N = 2p + |l|$ of these modes; rather, they facilitate a transformation between the two bases. In fact, the irreducible representation of the rotation under the LG basis is explicitly given by the Wigner D-matrix~\cite{biedenharn1984angular}, and the corresponding rotation angle is only determined by the difference between the Gouy phases along the two orthogonal directions~\cite{baladron2019isolating}, referred to as intermodal Gouy phase. This mechanism can also be adopted to explain the propagation behavior of STLG modes in isotropic dispersive media. The difference between spatial dispersion and temporal dispersion gives rise to that intermodal Gouy phase, which effectively determines the rotation angle in this SU(2) transformation. We will show that this rotation angle is governed by two key parameters of the STOVs: the spatiotemporal waist ratio ($\alpha=w_{\xi}/w_x$), and the GVD ($\beta_2$) of the media. This understanding further allows us to derive an analytical expression for the propagation of an arbitrary-order STLG mode.

\section{Intermodal Gouy phase}
Under the paraxial quasi-monochromatic approximation, the slowly varying amplitude $A$ of a pulse with central frequency $w_0$ propagating along the $z$-axis satisfies the following Schrödinger-type equation:
\begin{equation}
    \partial_x^2 A(\xi,x,z)-\beta_2 \partial_\xi^2 A(\xi,x,z)=2ik_0 \partial_z A(\xi,x,z).
    \label{eq1}
\end{equation}
where we have ignored the $y$-dependence for simplicity. The longitudinal accompanying spatiotemporal coordinate $\xi=z-v_g t$, the central wavenumber $ k_0=\omega_0 n_0/c$, and the media's refractive index, group velocity and GVD are denoted by $n_0$, $v_g$ and $\beta_2$, respectively. By employing the method of separation of variables and noticing the equivalence between diffraction in the transverse spatial direction and the spacetime direction  ${k_0\leftrightarrow k^\prime}_0=-k_0/\beta_2$, a complete set of propagation solutions can be immediately obtained, namely the STHG modes, which can be expressed as~\cite{hancock2021mode}
\begin{eqnarray}
\text{HG}^{\text{ST}}_{mn} && (x, \xi, z) = N_{mn} H_m \left[ \frac{\sqrt{2} \xi}{w_\xi(z)} \right] H_n\left[\frac{\sqrt{2} x}{w_x(z)} \right] \notag \\ && \times\text{exp}\left[- \frac{\xi^2}{w_\xi^2(z)} \right] \text{exp}\left[- \frac{x^2}{w_x^2(z)}\right]  \notag \\ && \times\text{exp}\left[ - \frac{ik_0 \xi^2}{2 \beta_2 R_\xi(z)} + \frac{i k_0 x^2}{2 R_x(z)} - i\psi_{mn}(z)\right] 
\label{eq2}
\end{eqnarray}
where $m$ and $n$ are the mode orders of the longitudinal spatiotemporal envelope and the transverse spatial envelope, with the waist sizes of pulse at $z=0$ specified by $w_{0\xi}$ and $w_{0x}$, respectively. Their corresponding Rayleigh lengths $z_{0\xi}=k_0w_{0\xi}^2/(2|\beta_2|)$ and $z_{0x}=k_0w_{0x}^2/2$, respectively. Then the waist sizes $w_{\xi,x}(z)=w_{0\xi,0x}\sqrt{1+z^2/z_{0\xi,0x}^2}$ at the propagation distance $z$, while the curvature radii of their quadratic wavefronts $R_{\xi,x}(z)=z+z_{0\xi,0x}^2/z$. The normalization constant $N_{mn}=\sqrt{2/[\pi 2^{m+n}m!n!w_\xi(z)w_x(z)]}$. The Gouy phase $\psi_{mn}(z)$ is sum of  two part: the longitudinal spatiotemporal Gouy phase $\psi_m(z)$ and transverse spatial Gouy phase $\psi_n(z)$, which depend on the propagation distance ( z ) and mode orders. This can be written as
\begin{align}
\psi_m(z) &= -\operatorname{sgn}(\beta_2)(m + \frac{1}{2}) \arctan\left( \frac{z}{z_{0\xi}} \right), \notag\\
\psi_n(z) &= (n + \frac{1}{2}) \arctan\left( \frac{z}{z_{0x}} \right).
\label{eq3}
\end{align}

We can nondimensionalize Eq. (\ref{eq2}) for further analysis when we do not consider the diffraction-induced overall broadening of the pulse size resulting from propagation. Firstly, the coordinates within the spatiotemporal $\xi-x$ plane are nondimensionalized as $[\mathrm{\Xi},X]=[\xi/w_\xi(z),x/w_x(z)]$. The propagation distances $z$ are normalized by the two Rayleigh distances into $Z_\xi=z/z_{0\xi}$ and $Z_x\equiv Z=z/z_{0x}$, satisfying a exchange relation of $Z_\xi/ Z_x=|\beta_2|\alpha^2$, where the waist ratio $\alpha=w_{0\xi}/w_{0x}$. Secondly, the quadratic phase terms along the longitudinal spatiotemporal and transverse spatial directions can be thereby written as
\begin{eqnarray}
 \frac{-k_0\xi^2}{2\beta_2R_\xi(z)}&&=-\text{sgn}(\beta_2)Z_\xi\mathrm{\Xi}^2,\notag\\ \frac{k_0x^2}{2R_x(z)}&&=Z_xX^2,  
 \label{eq4}
\end{eqnarray}
respectively. Therefore, the quadratic phase $s(\mathrm{\Xi},X)=-\text{sgn}(\beta_2)Z_\xi\mathrm{\Xi}^2+Z_xX^2$ changes its traditional circular entity to elliptical or even hyperbolic. In particular, the total Gouy phase term can also be rewritten as follows
\begin{eqnarray}
\psi_{mn}(z) && =\frac{N+1}{2}\left[\ -\text{sgn}\left(\beta_2\right)\arctan\left(Z_\xi\right)+\arctan{(Z_X)}\right] \notag \\ && \quad +\frac{l}{2}\left[ -\text{sgn}(\beta_2)\arctan{(Z_\xi)}-\arctan{(Z_X)}\right]\notag \\ && = \frac{N+1}{2}\sigma+\frac{l}{2}\delta
\label{eq5}
\end{eqnarray}
Here, we decompose the total Gouy phase into the first part $\left(N+1\right)\sigma/2$, which depends solely on the total order $N=m+n$ (denoted as extermodal Gouy phase), and the second part $l\delta/2$, which depends on the order difference $l=m-n$ (denoted as intermodal Gouy phase). Their dependencies on the propagation distance are represented as $\sigma=-\text{sgn}\left(\beta_2\right)\arctan\left(Z_\xi\right)+\arctan{(}Z_X)$ and $\delta=-\text{sgn}\left(\beta_2\right)\arctan\left(Z_\xi\right)-\arctan{(Z_X)}$. Finally, the expression of the STHG mode described by Eq. (2) can be reformulated in a dimensionless form as follows:
\begin{eqnarray}
    \text{HG}^{\text{ST}}_{mn} && (\mathrm{\Xi}, X,Z) =N_{mn}H_m(\sqrt2\mathrm{\Xi})H_n(\sqrt2X)e^{-\mathrm{\Xi}^2-X^2}\notag \\ && \times\exp {[{-i\frac{N+1}{2}\sigma (Z)-i\frac{l}{2}\delta (Z) +is(\Xi,X)]}}
    \label{eq6}
\end{eqnarray}
In the next section, we will elucidate, from the perspective of the action of the SU(2) group, that the intramodal Gouy phase plays a critical role in the evolution of the light intensity distribution of STLG modes during propagation in isotropic media, while the extermodal Gouy phase together with the quadratic phase only contribute a phase to the mode, do not govern the evolution of the light intensity distribution.

 \section{STLG modes and SU(2) symmetry}
The amplitude expression of the STLG mode with radial index $p$ and azimuthal index $l$ in the spatiotemporal plane at $z=0$ can be written as
 \begin{eqnarray}
     && \text{LG}^{\text{ST}}_{pl}(x,  \xi,0)=\text{LG}^{\text{ST}}_{pl}(R_{st},\mathrm{\Phi}_{st},0)\notag\\&&=N_{pl}(\sqrt2R_{st})^{|l|}L_p^{|l|}\left(2R_{st}^2\right)\exp({-R_{st}^2}+{il\mathrm{\Phi}_{st}})
     \label{eq7}
 \end{eqnarray}
Here, the transverse spatial width and the longitudinal temporal width of the pulse are denoted as $ w_{0x}$ and $w_{0\xi}$, respectively. The normalized radial coordinate is given by $R_{st}=\sqrt{\mathrm{\Xi}^2+X^2}=\sqrt{{\xi^2}/{w_{0\xi}^2}+{x^2}/{w_{0x}^2}}$ and the angular coordinate is defined as $\mathrm{\Phi}_{st}=\arctan({X}/{\mathrm{\Xi}})=\arctan({w_{0x}\xi}/{w_{0\xi}x})$. The normalization factor is $N_{pl}=\sqrt{2p!/[\pi(p+|l|)!w_{0x}w_{0\xi}]}$. In Appendix~\ref{sec:apA}, we derive the transformation relationship between the STHG modes and the STLG modes on the $z=0$  plane using SU(2) representation theory:
 \begin{eqnarray}
\text{LG}^{\text{ST}}_{Nl}\left(\mathrm{\Xi},X\right)=\sum_{\frac{l^\prime}{2}=-\frac{N}{2}}^{\frac{N}{2}}d_{\frac{l^\prime}{2},\frac{l}{2}}^{\frac{N}{2}}\left(-\frac{\pi}{2}\right)\text{HG}^{\text{ST}}_{Nl^\prime}\left(\mathrm{\Xi},X\right).
\label{eq8}
 \end{eqnarray}
Here, $d_{l^\prime/2,l/2}^{N/2}$ denotes the Wigner $d$-matrix, and the phase conventions for STLG and STHG modes are specified as:
 \begin{eqnarray}
     \text{LG}^{\text{ST}}_{Nl}(\mathrm{\Xi},X)&&={(-1)}^\frac{N-|l|}{2}\text{LG}^{\text{ST}}_{pl}(\mathrm{\Xi},X),\notag\\ \text{HG}^{\text{ST}}_{Nl}(\mathrm{\Xi},X)&&={(-i)}^\frac{N-l}{2}\text{HG}^{\text{ST}}_{mn}(\mathrm{\Xi},X).\\ \notag
     \label{eq9}
 \end{eqnarray}
Using Eqs. (\ref{eq6}) and (\ref{eq8}), the STLG mode at $z=Zz_{0x}$  becomes
\begin{eqnarray}
    \text{LG}^{\text{ST}}_{Nl}(\mathrm{\Xi},&& X,Z)=\sum_{\frac{l^\prime}{2}=-\frac{N}{2}}^{\frac{N}{2}}d_{\frac{l^\prime}{2},\frac{l}{2}}^{\frac{N}{2}}(-\frac{\pi}{2})\text{HG}^{\text{ST}}_{Nl^\prime}(\mathrm{\Xi},X,Z) \notag\\
    &&=\sum_{\frac{l^\prime}{2}=-\frac{N}{2}}^{\frac{N}{2}}\exp({-i\frac{l^\prime}{2}\delta})d_{\frac{l^\prime}{2},\frac{l}{2}}^{\frac{N}{2}}(-\frac{\pi}{2})\text{HG}^{\text{ST}}_{Nl^\prime}(\mathrm{\Xi},X) \notag\\
    &&\quad\times \exp({-i\frac{N+1}{2}\sigma}+is).
    \label{eq10}
\end{eqnarray}
Here, the intermodal Gouy phase depends on the summation index $l^\prime$, so it must be kept inside the summation, whereas the extermodal Gouy phase depends only on the total order $N$, and can be factored out. According to Appendix~\ref{sec:apA}, the STHG mode at $z =0$ can also be expressed as a linear combination of STLG modes via the inverse of Eq. (\ref{eq8}). Thus, Equation (\ref{eq10}) yields
\begin{widetext}
 \begin{eqnarray}
\text{LG}^{\text{ST}}_{Nl}(\mathrm{\Xi},X,Z)&&=\exp({is-i\frac{N+1}{2}\sigma})\sum_{\frac{l^\prime}{2}=-\frac{N}{2}}^{\frac{N}{2}}\sum_{\frac{l^{\prime\prime}}{2}=-\frac{N}{2}}^{\frac{N}{2}}{d_{\frac{l^\prime}{2},\frac{l}{2}}^{\frac{N}{2}}(-\frac{\pi}{2})e^{-il^\prime\delta/2}d_{\frac{l^{\prime\prime}}{2},\frac{l^\prime}{2}}^{\frac{N}{2}}(\frac{\pi}{2})\text{LG}^{\text{ST}}_{Nl^{\prime\prime}}(\mathrm{\Xi},X)}\notag\\&&=\exp({is-i\frac{N+1}{2}\sigma})\exp(-i\frac{\pi}{4}{\hat{Q}}_2)\exp(-i\frac{\delta}{2}{\hat{Q}}_3)\exp(i\frac{\pi}{4}{\hat{Q}}_2)\text{LG}^{\text{ST}}_{Nl} (\mathrm{\Xi},X).
\label{eq11}
\end{eqnarray} 
By exploiting the $su(2)$ algebra satisfied by the conserved quantities ${\hat{Q}}_1, {\hat{Q}}_2$ and ${\hat{Q}}_3$ (defined in Appendix~\ref{sec:apA}), we obtain the identity:
\begin{eqnarray}
    &&\exp(-i\frac{\pi}{4}{\hat{Q}}_2)\exp(-i\frac{\delta}{2}{\hat{Q}}_3)\exp(i\frac{\pi}{4}{\hat{Q}}_2)=\exp(-i\frac{\delta}{2}{\hat{Q}}_1)=\hat{D}(-\frac{\pi}{2},\delta,\frac{\pi}{2})\in \text{SU}(2),
    \label{eq12}
\end{eqnarray}
\end{widetext}
\begin{figure*}
    \centering
    \includegraphics[width=16cm]{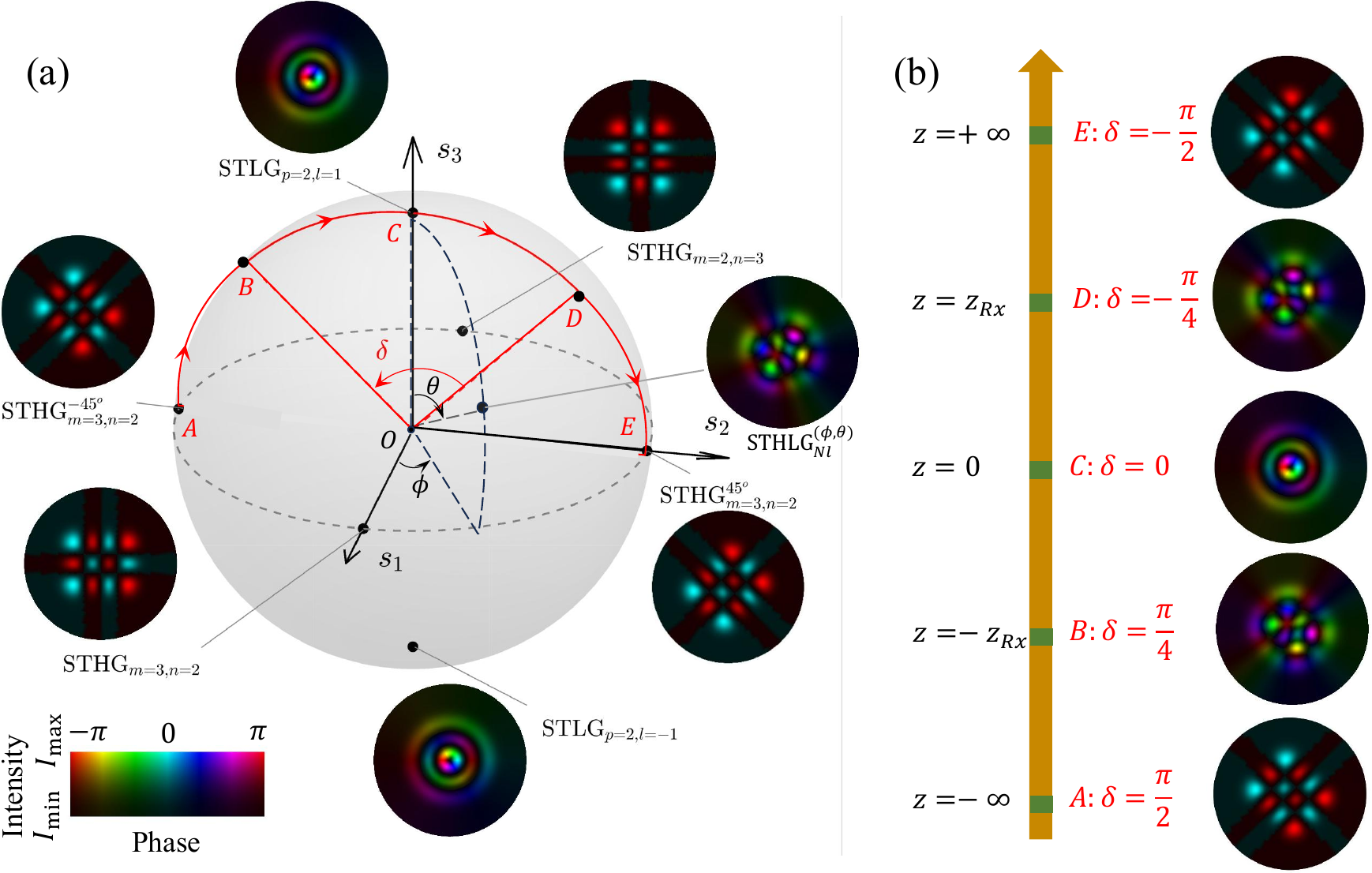}
    \caption{\label{Fig_1}(a) The STMPS with order $N=5$ and $l=1$. Every point on the sphere represents a spatiotemporal Gaussian mode and the evolution of the modes propagating in the isotropic media can be described as the rotation around the $s_1$-axis with the rotation angle the same as the intermodal Gouy phase $\delta$. (b) For the free space case, the change of $\delta$ with the propagation distance $z$: $\pi/2\rightarrow0\rightarrow-\pi/2$ induce the change of mode: $HG_{mn}^{ST,{-45}^o}\rightarrow LG_{pl}^{ST}\rightarrow HG_{mn}^{ST,+{45}^o}$. The images $A-E$ of (b) correspond to the points $A-E$ of the red curve in (a). }
\end{figure*}
where $\hat{D}(\alpha,\beta,\gamma)$ denotes the SU(2) transformation parameterized by Euler angles $\{\alpha,\beta,\gamma\}$. Since the phase factor $s(\mathrm{\Xi},X)+(N+1)\sigma(Z)/2$ is independent of the modal indices and does not affect the intensity distribution, we define the normalized STLG amplitude as $LG^{ST}_{Nl}$, to distinguish it from the unnormalized $\text{LG}_{Nl}^{\text{ST}}$. Its propagation can then be interpreted as a rotation $\exp(-i\frac{\delta}{2}{\hat{Q}}_1)$ generated by ${\hat{Q}}_1$ acting on the input STLG mode:
\begin{equation}
    {LG}^{ST}_{Nl} \left(\mathrm{\Xi},X,Z\right)=\sum_{\frac{l^\prime}{2}=-\frac{N}{2}}^{\frac{N}{2}}e^{-il^\prime\frac{\pi}{4}}{d_{\frac{l^\prime}{2},\frac{l}{2}}^{\frac{N}{2}}(\delta)}e^{il\frac{\pi}{4}}LG^{ST}_{Nl^\prime} (\Xi,X).
    \label{eq13}
\end{equation}
This expression represents the analytical solution for the propagation of an arbitrary STLG mode with radial index $p=(N-|l|)/2$ and azimuthal index $l$ in an isotropic media. It clearly shows that the LG spectrum spreads within the degenerate subspace during propagation, which is contrast to the self-similar transformation behavior of purely spatial LG modes~\cite{alonso2017ray}, that is $LG_{Nl}(X,Y,Z)\propto LG_{Nl}(X,Y,0)$.

Analogous to the modal Poincaré sphere (MPS) in transverse spatial dimensions~\cite{shen20202}, we can construct a spatiotemporal modal Poincaré sphere (STMPS). The STMPS is defined as a unit sphere embedded in a three-dimensional Cartesian coordinate system $(s_1,s_2,s_3)$, where the north pole represents the $LG^{ST}_{Nl}$ mode. The point on the sphere specified by polar angle $\theta$ and azimuthal angle $\phi$ represents the general spatiotemporal Laguerre-Hermite-Gaussian (STHLG) mode~\cite{abramochkin2004generalized}, defined as $HLG_{Nl}^{ST,(\phi,\theta)}(\Xi,X)=\hat{D}(\phi,\theta,0)LG_{Nl}^{ST}(\Xi,X)$, where STHG modes lie on the equator of the STMPS. The STMPS with $N=5$ and $l=1$ is displayed in Fig.~\ref{Fig_1}, where some special points on the sphere are mapped with corresponding patterns of phase and intensity representative of the respective spatiotemporal modes. Subsequently, we establish a correspondence between the SO(3) rotation on the STMPS and SU(2) transformation of STHLG modes: A rotation around the $s_3$-axis by  an angle $\phi$ corresponds to a visual rotation of the beam pattern by an angle $\phi/2$, which is mathematically described by the operator $\exp(-i\frac{\phi}{2} \hat{Q}_3)$. A rotation around the $s_2$-axis by angle $\theta$, represented by $\exp(-i\frac{\theta}{2} \hat{Q}_2)$, corresponds to a gyration transformation—an operation that does not manifest as a simple visual rotation. A rotation around the $s_1$-axis by angle $\delta$, implemented by $\exp(-i\frac{\delta}{2} \hat{Q}_1)$, corresponds to an asymmetric fractional Fourier transform. 
\begin{figure*}
    \centering
    \includegraphics[width=16cm]{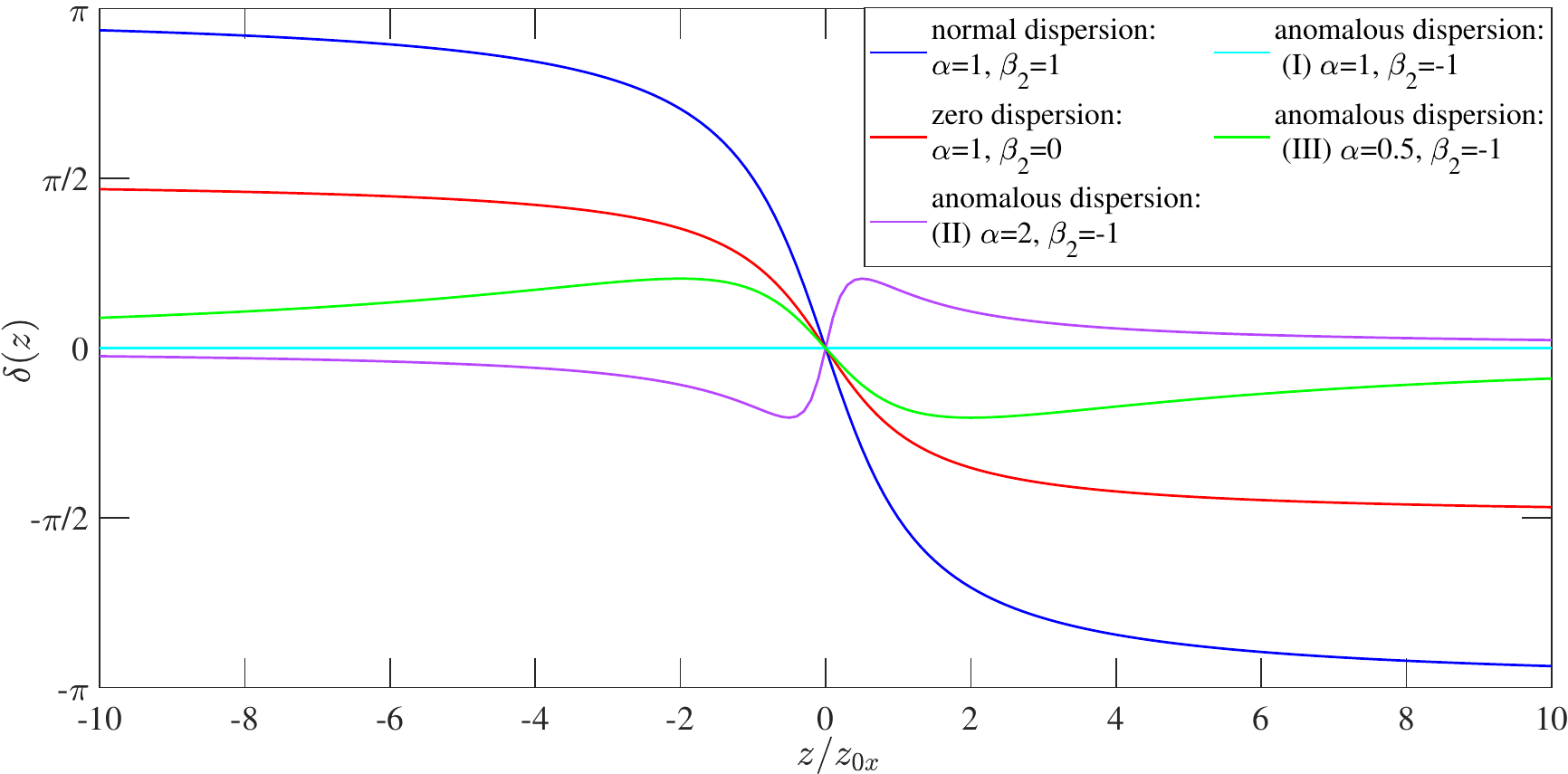}
    \caption{\label{Fig_2}The intermodal Gouy phase $\delta$ as a function of the propagation distance $z$ for several different pulse elliptically $\alpha$ and GVD $\beta_2$. All the curves can be categorized into three classes, which have distinguished convergence and monotonic behavior, corresponding to zero dispersion, normal dispersion and anomalous dispersion, respectively.}
\end{figure*}

Above all, the propagation of STHLG modes in the isotropic media can be intuitively interpreted as a rotation of the corresponding point on the STMPS around the $s_1$-axis. The rotation angle is precisely equivalent to the intermodal Gouy phase ($\delta$), with this rotational motion tracing a red trajectory on the sphere depicted in Fig.~\ref{Fig_1}(a). For instance, the complete propagation process in free space can be elucidated as follows: the initial point located on the equator (the $-45^o$ tilted STHG mode at $z=-\infty$) undergoes a clockwise rotation of $\pi/2$ to reach the north pole (the STLG mode at $z = 0$). This rotation continues by an additional $\pi/2$, culminating in the final point located on the equator (the $+45^o$ tilted STHG mode at $z =+\infty$), as shown in Fig.~\ref{Fig_1}(b). Specifically, the non-tilted STHG mode, lying on the $s_1$-axis, remains invariant in its intensity distribution during propagation. Furthermore, considering the Wigner distribution function associated with the spatiotemporal mode $E(\mathrm{\Xi},X)$ in phase space, we have
\begin{eqnarray}
  W(\boldsymbol{V})=\iint && { E(\mathrm{\Xi}+\frac{\mathrm{\Xi}^\prime}{2},X+\frac{X^\prime}{2})E^\ast(\mathrm{\Xi}-\frac{\mathrm{\Xi}^\prime}{2},X-\frac{X^\prime}{2})}\notag\\&&\times \exp({-iK_\mathrm{\Xi}\mathrm{\Xi}^\prime-iK_XX^\prime})d\mathrm{\Xi}^\prime d X^\prime,
  \label{eq14}
\end{eqnarray}
where the position in phase space is denoted as ${\boldsymbol{V}}=(\mathrm{\Xi},K_\mathrm{\Xi},X,K_X)$. Then the action of the operator $\exp(-i\frac{\delta}{2}{\hat{Q}}_1)$, corresponding to a rotation around the $s_1$-axis, induces a visual rotation of the Wigner function in phase space. Denoting the transformed Wigner function as $W^\prime(\boldsymbol{V})$, we have $W^\prime\left(\boldsymbol{V}\right)=W\left(S^{-1}\boldsymbol{V}\right)$, where S is the $4\times4$ rotation matrix:
\begin{equation}
    S=\left[\begin{matrix}\cos{\frac{\delta}{2}}&\sin{\frac{\delta}{2}}&0&0\\-\sin{\frac{\delta}{2}}&\cos{\frac{\delta}{2}}&0&0\\0&0&\cos{\frac{\delta}{2}}&-\sin{\frac{\delta}{2}}\\0&0&\sin{\frac{\delta}{2}}&\cos{\frac{\delta}{2}}\\\end{matrix}\right].
    \label{eq15}
\end{equation}
This implies that the pattern undergoes a rotation by angle $-\frac{\delta}{2}$ on the $(\mathrm{\Xi},K_\mathrm{\Xi})$ plane and simultaneously by angle $\frac{\delta}{2}$ on the $(X,\ K_X)$ plane.

\section{Interpretation of evolution for different dispersion media}
From the above analysis, we find that the influence of propagation on the intensity distribution of STLG modes is uniquely determined by the intermodal Gouy phase $\delta$ (i.e., the rotation angle). According to Eq. (\ref{eq5}), $\delta $ depends not only on the GVD of the media ($\beta_2$), and the pulse ellipticity ($\alpha$), but also on the propagation distance ($Z$). The explicit functional form is given by
\begin{equation}
    \delta\left(\alpha,\beta_2,Z\right)=-\arctan{\left(\beta_2\alpha^2Z\right)-\arctan{Z}}.
    \label{eq16}
\end{equation}
This expression consists of the sum of two arctangent-type functions, both of which vanish at $z =0$. The asymptotic behavior as $z\rightarrow\pm\infty$ depends on the dispersion type of the media. Figure~\ref{Fig_2} shows the evolution of $\delta$ with respect to $z$ for various values of $\alpha$ and $\beta_2$. The curves fall into three distinct classes corresponding to zero dispersive, normal dispersive, and anomalous dispersive media. We will provide a detailed analysis of each category in the following discussions.
\begin{figure*}
    \centering
    \includegraphics[width=16cm]{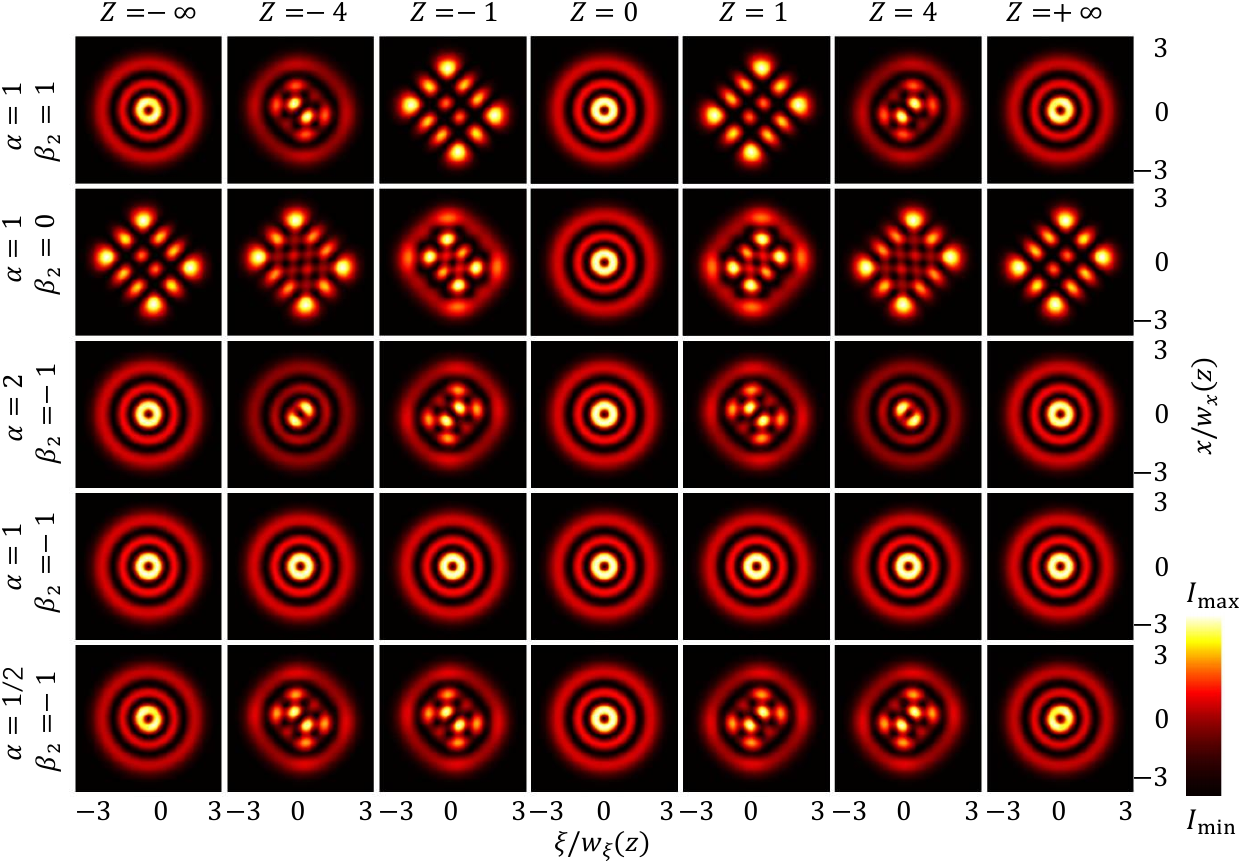}
    \caption{\label{Fig_3}The evolution of the intensity distribution of the STLG modes propagating in an isotropic media under various combinations of pulse ellipticity $\alpha$ and group velocity dispersion $\beta_2$. The underlying mechanism is interpreted as a rotation on the STMPS around the $s_1$-axis (see Fig.~\ref{Fig_1}), driven by the intermodal Gouy phase shown in Fig.~\ref{Fig_2}. The five distinct conditions corresponding to different parameter settings are analyzed in detail in the main text.}
\end{figure*}

{\bf{Case 1}}: Zero dispersion (${\beta}_2 = 0$): For a zero dispersive media, such as free space, the first term on the right-hand side of Eq.~(\ref{eq16}) vanishes regardless of the ellipticity $\alpha$. Thus, $\delta$ reduces to the second term, $\delta=-\arctan{Z}$, which monotonically decreases from $\pi/2$ to $0$ and then to $-\pi/2$ as $z$ increases from $-\infty$ to $0$ and then to $+\infty$, as shown by the red curve in Fig.~\ref{Fig_2}. On the STMPS, this corresponds to a clockwise rotation around the $s_1$-axis from the equatorial point at ${45}^\circ$, through the north pole, and finally to the equatorial point at$ -{45}^\circ$. This motion delineates a half-circle trajectory (illustrated by the red arc in Fig.~\ref{Fig_1}(a)), resulting in an overall rotation angle of $-\pi$. The mode evolution can be described as follows: a far-field $HG_{N,l}^{ST,-45^\circ}$ mode is  transformed into a near-field $LG^{ST}_{N,l}$ mode, and then evolves into  a far-field $HG_{N,l}^{ST,+{45}^\circ}$ mode, as shown in the second row of Fig.~\ref{Fig_3}. When including the quadratic phase and the extermodal Gouy phase, the resulting field distribution agrees with previous works~\cite{porras2023propagation,hyde2023propagation}.

{\bf{Case 2}}:  Normal dispersion (${\beta}_2 > 0$): In the case of a normal dispersive media, both terms in Eq. (\ref{eq16}) are arctangent-type functions with the same sign. The ellipticity $\alpha$ controls the convergence rate of the curve: a larger $\alpha$ leads to a sharper rotation near the origin, and hence a steeper slope. Over the full propagation range, $\delta$ varies from $\pi$ to 0 and then to $-\pi$ as $z$ goes from $-\infty$ to 0 and then to$+\infty$. This behavior results in an effective doubling of the total variation when compared to the zero dispersive scenario, as illustrated by the blue monotonic curve displayed in Fig.~\ref{Fig_2}. On the STMPS, this corresponds to a full great-circle rotation around the $s_1$-axis, starting at the south pole, passing through the equator at ${45}^\circ$, reaching the north pole, and then returning to the south pole. The mode evolution is as follows: In  the negative half-space, a far-field $LG^{ST}_{N,-l}$ mode with opposite topological charge transforms into an $HG^{ST,-{45}^\circ}_{N,l}$ mode and then evolves into a near-field $LG^{ST}_{N,l}$ mode; In the positive half-space, the process reverses, returning to the initial $LG^{ST}_{N,-l}$ mode, as shown in the second row of Fig.~\ref{Fig_3}.

{\bf{Case 3}}:  Anomalous dispersion (${\beta}_2< 0$): For anomalous dispersive media, the two terms in Eq. (\ref{eq16}) have opposite signs. This causes them to cancel out at infinity, resulting in $\delta(\pm\infty)\ =\ 0$. Furthermore, since $\delta(0)\ =\ 0$, it follows that the function $\delta(z)$ is non-monotonic. The behavior in the intermediate regime depends further on the value of the ellipticity $\alpha$:

(I). When $\alpha=1/\sqrt{-\beta_2}$, the two terms cancel exactly over the full propagation range, yielding $\delta\left(z\right)\equiv0$, as shown by the cyan curve in Fig.~\ref{Fig_2}. On the STMPS, the state remains fixed at the north pole. In this case, the STLG mode's intensity distribution remains invariant, exhibiting a self-similar structure akin to the evolution of a 2D transverse Gaussian mode, as illustrated in the fourth row of Fig.~\ref{Fig_3}.

(II). When $\alpha\neq1/\sqrt{-\beta_2}$, the $\delta(z)$ curve attains extreme values at $Z=\pm 1 / (\alpha \sqrt{-\beta_2})$, given by $\delta_{\text{ext}}=\pm \arctan[(\alpha\sqrt{-\beta_2}-1/(\alpha\sqrt{-\beta_2}))/2]$.  For two symmetric points $Z_1$ and $Z_2$ satisfying $ Z_1 Z_2=1/[\alpha^2(-\beta_2)]$, the intermodal Gouy phase is equal: $\delta(Z_1) =\delta(Z_2)$. When $\alpha > 1/\sqrt{-\beta_2}$, the first term predominates over the second in the near field, thus indicating that $\delta(z)$ is predominantly influenced by this term. In this case, $\delta$ is negative in the negative half-space and positive in the positive half-space, resulting in an odd-symmetric function, as shown by the magenta curve in Fig.~\ref{Fig_2}. On the STMPS, the trajectory starts at the north pole, rotates clockwise to the minimal angle $\delta_{\text{min}}$, reverses back counterclockwise to the north pole, continues counterclockwise to the maximum angle $\delta_{\text{max}}$, and finally returns to the north pole clockwise. This evolution corresponds to the mode sequence: a far-field $LG_{N,l}^{ST}$ mode transforms into an $HLG_{N,l}^{ST,\left(-\pi/2,\delta_{\text{min}}\right)}$ mode, then into a near-field $STLG_{N,l}$ mode, further into an $HLG_{N,l}^{ST,\left(\pi/2,\delta_{\text{max}}\right)}$ mode, and finally back into the far-field $LG_{N,l}^{ST}$ mode, as shown in the third row of Fig.~\ref{Fig_3}. Notably, all intermediate stages (except for the maximal STHLG mode) involve repeated visits to the same STLG mode, indicating a cycle of mode destruction and reconstruction, reminiscent of Talbot effect~\cite{wen2013talbot}. 

(III). When $\alpha<1/\sqrt{-\beta_2}$, the entire rotation process is reversed compared to the $\alpha>1/\sqrt{-\beta_2}$ case. Accordingly, both the evolution of $\delta(z)$ and the intensity distribution of the modes evolve in the opposite direction, as depicted by the green curve in Fig.~\ref{Fig_2} and the fifth row of Fig.~\ref{Fig_3}.

\section{Conclusion}
By analyzing the propagation equation of spatiotemporal Gaussian wave packets in isotropic dispersive media, we reveal the underlying SU(2) symmetry inherent in spatiotemporal Gaussian modes and establish the concept of the STMPS. The evolution of STLG modes can be interpreted as a rotation on the STMPS around the $s_1$-axis, with the rotation angle precisely corresponding to the intermodal Gouy phase. Based on the theory of irreducible representations of the SU(2) group, we derive the general propagation expressions for complete STLG modes with arbitrary radial and azimuthal indices in isotropic dispersive media. In the special case of $p=0$, our results reduce to those reported in Ref.~\cite{hyde2023propagation} (see Appendix~\ref{sec:apB} for details). Beyond the STLG modes, any mode in the $N+1$-th order degenerate subspace can be described by the Husimi function—representing the associated Majorana constellation~\cite{gutierrez2020modal}, and its rotation on the coherent-state sphere around the $s_1$-axis can be used to interpret its propagating evolution, which will be presented in our forthcoming work. The dependence of  the intermodal Gouy phase on propagation distance, determined by the GVD and ellipticity, can be classified into three distinct regimes: normal dispersion, zero dispersion, and anomalous dispersion . Each regime corresponds to a qualitatively different trajectory on the STMPS.  Notably, in anomalous dispersion media, the non-monotonic dependence leads to disruptions and recurrences of the modal intensity profile, akin to the Talbot effect. Our findings provide a symmetry-based framework for understanding the propagation dynamics of spatiotemporal vortices in dispersive media, which holds promising applications for the manipulation of spatiotemporal light fields. 

\begin{acknowledgments}
This work was supported by the National Natural Science  Foundation of China (NSFC, Grant No. 12265004), Jiangxi Provincial Natural Science Foundation (Grant No. 20242BAB26010), and Ph.D.
Research Foundation (BSJJ202438).
\end{acknowledgments}

\appendix

\section{\label{sec:apA}Two-dimensional spatiotemporal harmonic oscillator}

Analogous to the isotropic harmonic oscillator in two-dimensional spatial domain, the Hamiltonian of oscillator in the spatiotemporal domain can be defined by
\begin{equation}
    \hat{H}=\frac{1}{2}{(\mathrm{\Xi}}^2+X^2-\partial_\mathrm{\Xi}^2-\partial_X^2).
    \label{eqA1}
\end{equation}
We define the creation and annihilation operators along the $\mathrm{\Xi}$ and X directions as
\begin{align}
    {\hat{a}}_\mathrm{\Xi}^\dag & =\left(\mathrm{\Xi}-\partial_\mathrm{\Xi}\right)/\sqrt2, \notag\\ {\hat{a}}_\mathrm{\Xi}&=\left(\mathrm{\Xi}+\partial_\mathrm{\Xi}\right)/\sqrt2,\notag\\{\hat{a}}_X^\dag & =(X-\partial_X)/\sqrt2,\notag\\  {\hat{a}}_X&=(X+\partial_X)/\sqrt2,
    \label{eqA2}
\end{align}
which satisfy the commutation relations $[{\hat{a}}_i^\dag,\ {\hat{a}}_j]=\delta_{ij}; i,j\in \{\Xi,X\}$. The Hamiltonian can then be rewritten as $\hat{H}\equiv{\hat{Q}}_0={\hat{a}}_\mathrm{\Xi}^\dag{\hat{a}}_\mathrm{\Xi}+{\hat{a}}_X^\dag{\hat{a}}_X+1$. There exist three conserved quantities expressed as
\begin{eqnarray}
&&\hat{Q}_1={\hat{a}}_\mathrm{\Xi}^\dag{\hat{a}}_\mathrm{\Xi}-{\hat{a}}_X^\dag{\hat{a}}_X,\notag\\ 
&& {\hat{Q}}_2={\hat{a}}_\mathrm{\Xi}^\dag{\hat{a}}_X+{\hat{a}}_X^\dag{\hat{a}}_\mathrm{\Xi},\notag\\
&& {\hat{Q}}_3=i({\hat{a}}_X^\dag{\hat{a}}_\mathrm{\Xi}-{\hat{a}}_\mathrm{\Xi}^\dag{\hat{a}}_X),
\label{eqA3}
\end{eqnarray}
which obey the $su(2)$ algebra $[{\hat{Q}}_i,{\hat{Q}}_j]=i\varepsilon_{ijk}\hat{Q}_k$, where $i,j,k=1,2,3$ and $\varepsilon_{ijk}$ is the Levi-Civita symbol. The simultaneous eigenstates of ${\hat{Q}}_0$ and ${\hat{Q}}_3$ are the STLG modes:
\begin{eqnarray}
\hat{Q}_0LG^{ST}_{pl}\left(\mathrm{\Xi},X\right)&&=\left(N+1\right)LG^{ST}_{pl}\left(\mathrm{\Xi},X\right),\notag\\
{\hat{Q}}_3LG^{ST}_{pl}(\mathrm{\Xi},X)&&=lLG^{ST}_{pl}(\mathrm{\Xi},X);
\label{eqA4}
\end{eqnarray}
and the simultaneous eigenstates of ${\hat{Q}}_0$ and ${\hat{Q}}_1$ are the STHG modes:
\begin{eqnarray}
    {\hat{Q}}_0HG^{ST}_{mn}(\mathrm{\Xi},X)&&=(N+1)HG^{ST}_{mn}(\mathrm{\Xi},X),\notag\\{\hat{Q}}_1HG^{ST}_{mn}(\mathrm{\Xi},X)&&=(m-n)HG^{ST}_{mn}(\mathrm{\Xi},X);
    \label{eqA5}
\end{eqnarray}
\begin{figure*}
    \centering
    \includegraphics[width=14cm]{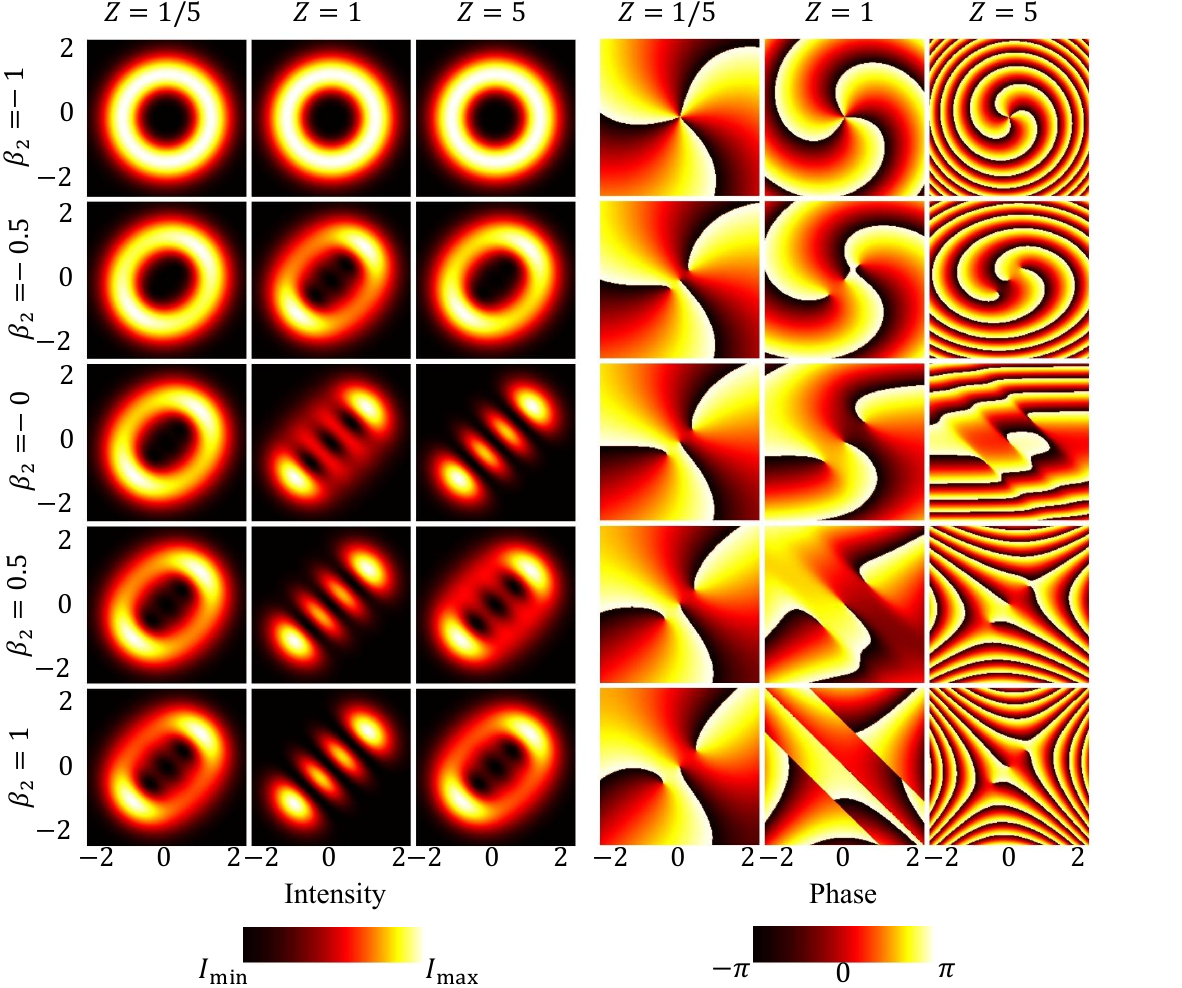}
    \caption{\label{Fig_4}The  evolutions of the intensity (left side) and phase distribution (right side) of the circular type ($\alpha=1$) STLG mode with $p=0$ and $l=3$ propagating in the media with several GVD: $\beta_2=-1,\ -0.5,\ 0,\ 0.5,1$. revisiting the result given Fig. 1 in Ref. ~\cite{hyde2023propagation}.}
\end{figure*}
Here, the total mode order is $N=2p+|l|=m+n$, and the energy eigenvalue is $N+1$. 
\begin{widetext}
The eigenspace associated with each $N$ is $N+1$-dimensional and is in variant under the action of the conserved quantities ${\hat{Q}}_1,{\hat{Q}}_2,{\hat{Q}}_3$, which generate the SU(2) group. Thus, each degenerate eigenspace forms a representation of SU(2). To facilitate the use of SU(2) representation theory in relating the STLG and STHG modes, we introduce the following phase and index conventions:
\begin{equation}
    LG^{ST}_{Nl}(\mathrm{\Xi},X)={(-1)}^\frac{N-|l|}{2}LG^{ST}_{pl}(\mathrm{\Xi},X), \quad \quad HG^{ST}_{Nl}(\mathrm{\Xi},X)={(-i)}^\frac{N-l}{2}HG^{ST}_{mn}(\mathrm{\Xi},X),
    \label{eqA6}
\end{equation}
where $l=m-n$ and the added overall phase is known as the Condon–Shortley phase. 
In fact, this SU(2) representation is actually irreducible, on which the group action of linear space is well known. When the group elements of SU(2) are parameterized by Euler angles $\{ \alpha,\beta,\gamma\}$ as
\begin{equation}
    \hat{D}(\alpha,\beta,\gamma)=\exp(-i\frac{\alpha}{2}{\hat{Q}}_3)\exp(-i\frac{\beta}{2}{\hat{Q}}_2)\exp(-i\frac{\gamma}{2}{\hat{Q}}_3),
    \label{eqA7}
\end{equation}
and taking the STLG modes  as a basis for the $N+1$-th subspace, i.e., $\{ \ LG^{ST}_{Nl}(\mathrm{\Xi}, X) \ | \ l=-N,-N+2,\cdots,N-2,N\}$, the action of $\hat{D}(\alpha,\beta,\gamma)$ takes following matrix form:
\begin{equation}
 \hat{D}(\alpha,\beta,\gamma)LG^{ST}_{Nl}(\mathrm{\Xi},X)= \sum_{\frac{l^\prime}{2}=-\frac{N}{2}}^{\frac{N}{2}}{D_{\frac{l}{2},\frac{l^\prime}{2}}^{\frac{N}{2}}(\alpha,\beta,\gamma)}LG^{ST}_{Nl^\prime}(\mathrm{\Xi},X), 
\end{equation}
where $D^{N/2}$ is the Wigner $D$-matrix given by
\begin{equation}
    D_{\frac{l}{2},\frac{l^\prime}{2}}^{\frac{N}{2}}(\alpha,\beta,\gamma)=\exp{(-i\frac{l^\prime}{2}\alpha)}d_{\frac{l}{2},\frac{l^\prime}{2}}^{\frac{N}{2}}\left(\beta\right)\exp{(-i\frac{l}{2}\gamma)},
    \label{eqA8}
\end{equation}
and the Wigner $d$-matrix is defined as ~\cite{wigner2012group}
\begin{equation} 
   d_{m,m^\prime}^j(\beta)=
   \sum_{v=min(j+m,j-m^\prime)}^{max(0,m-m^\prime)}\frac{{(-1)}^{m^\prime-m+v}\sqrt{(j+m)!(j-m)!(j+m^\prime)!(j-m^\prime)!}}{(j+m-v)!(j-m^\prime+v)!(m^\prime-m+v)!v!}(\cos\frac{\beta}{2})^{2j+m-m^\prime-2v}(\sin\frac{\beta}{2})^{m^\prime-m+2v}.
   \label{eqA9}
\end{equation}
Therefore, the STLG mode is actually the transformation of the STHG mode by $\hat{D}(0,-\frac{\pi}{2},0)$:
\begin{equation}
    LG^{ST}_{Nl}(\mathrm{\Xi},X)=\sum_{\frac{l^\prime}{2}=-\frac{N}{2}}^{\frac{N}{2}}d_{\frac{l^\prime}{2},\frac{l}{2}}^{\frac{N}{2}}(-\frac{\pi}{2})HG^{ST}_{Nl^\prime}(\mathrm{\Xi},X).
\end{equation}
Similarly, a generalized STHLG mode with parameters $(\theta,\phi)$ is related to the STLG mode via
\begin{equation}
    { HLG}_{Nl}^{ST,(\theta,\phi)}(\mathrm{\Xi},X)=\hat{D}(\phi,\theta,0)LG^{ST}_{Nl}(\mathrm{\Xi},X)=\sum_{\frac{l^\prime}{2}=-\frac{N}{2}}^{\frac{N}{2}}d_{\frac{l^\prime}{2},\frac{l}{2}}^{\frac{N}{2}}(\theta)e^{i\frac{l^\prime}{2}\phi}LG^{ST}_{Nl^\prime}(\mathrm{\Xi},X).
    \label{eqA11}
\end{equation}
\end{widetext}
\section{\label{sec:apB}Revisiting the result for the STLG mode with ${p}=\mathbf{0}$ and ${l}=\mathbf{3}$}
In Ref.~\cite{hyde2023propagation}, the authors derived the propagation expression for a special case of the STLG mode with $p=0$ and $l=3$ in a dispersive media, which agree well with the precise numerical results obtained by using the FFT algorithm. Here, using our generalized derivation approach, we are able to revisit the result (as shown in Fig.~\ref{Fig_4}) presented in Fig. 1 in Ref.~\cite{hyde2023propagation}. It is important to note that the evolution of the intensity distribution is solely governed by the rotation around the $s_1$-axis, whereas the evolution of the phase distribution involves not only the mode rotation, but also the contributions from the quadratic phase $s$ and the extramodal Gouy phase $\sigma$.

\nocite{*}

\bibliography{SU_of_STLG_modes.bib}

\end{document}